\documentclass[preprint,12pt]{elsarticle}
\usepackage{amssymb}
\usepackage{amsthm}
\usepackage{amsmath}
\usepackage{adjustbox}

\usepackage{multirow}
\usepackage{tabularx}
\usepackage{booktabs}
\journal{Acta Materialia}

\begin{document}

\begin{frontmatter}


% \title{Ultrasonic determination of crystallographic texture regardless of medium dispersiveness\tnoteref{label1}}
% \title{Simultaneous determination of crystallographic texture and elastic constants with ultrasonic goniometry\tnoteref{label1}}
\title{Ultrasonic determination of crystallographic texture by transmitted field fitting regardless of medium dispersivity.}
% \tnotetext[label1]{nota titulo1}

\author[a,b]{Diego A. Cowes}
\ead{diegocowes@cnea.gob.ar}
% \fntext[label2]{nota autor}

\author[a,b,c]{Martín P. Gómez}
\author[a,b]{Ignacio Mieza}
%%\cortext[cor1]{hola}

\affiliation[a]{organization={Comisión Nacional de Energía Atómica},
            addressline={Gral. Paz 1499},
            city={Villa Maipú},
            postcode={B1650},
            state={Bs. As.},
             country={Argentina}}
\affiliation[b]{organization={Instituto Sabato},
            addressline={Gral. Paz 1499},
            city={Villa Maipú},
            postcode={B1650},
            state={Bs. As.},
             country={Argentina}}
\affiliation[c]{organization={Universidad Tecnológica Nacional},
            addressline={San Martín 1171},
            city={Campana},
            postcode={B2804},
            state={Bs. As.},
             country={Argentina}}


\begin{abstract}
The determination of crystallographic texture through elastic wave propagation offers a cost-effective, nondestructive means of obtaining through-thickness information with minimal sample preparation. Existing ultrasonic approaches rely on either bulk-wave or guided-wave velocity measurements for texture inversion. These strategies impose geometric constraints: bulk-wave methods become impractical for thin specimens, whereas guided-wave techniques are limited to relatively small thicknesses. Furthermore, many formulations assume orthotropic symmetry of the aggregate, thereby restricting their applicability to materials with higher anisotropy. In this work, a full-field wave fitting strategy is developed in which the transmitted ultrasonic field is simulated and directly compared to experimental measurements. Because the approach does not rely on bulk-wave or plate-wave approximations, it remains applicable across a broad range of specimen thicknesses. Furthermore, no macroscopic symmetry assumptions are imposed on the aggregate, enabling the characterization of generally anisotropic materials. The effective elastic response is computed using a Hashin–Shtrikman homogenization framework, which provides tighter bounds than classical Voigt–Reuss–Hill averages and constrains the admissible search space during optimization, thereby improving convergence. The nonlinear inverse problem is solved using a GPU-accelerated optimization scheme. The methodology is validated on materials with hexagonal and cubic crystal symmetry over a range of specimen thicknesses. The inferred texture coefficients show consistent agreement with independent diffraction measurements. Additionally, textures with weak elastic anisotropy are successfully recovered, demonstrating the robustness and versatility of the proposed method. Complete measurement and inversion are achieved within approximately 10 minutes.
\end{abstract}

\begin{graphicalabstract}
\vfill
\centering
\includegraphics[width=15cm]{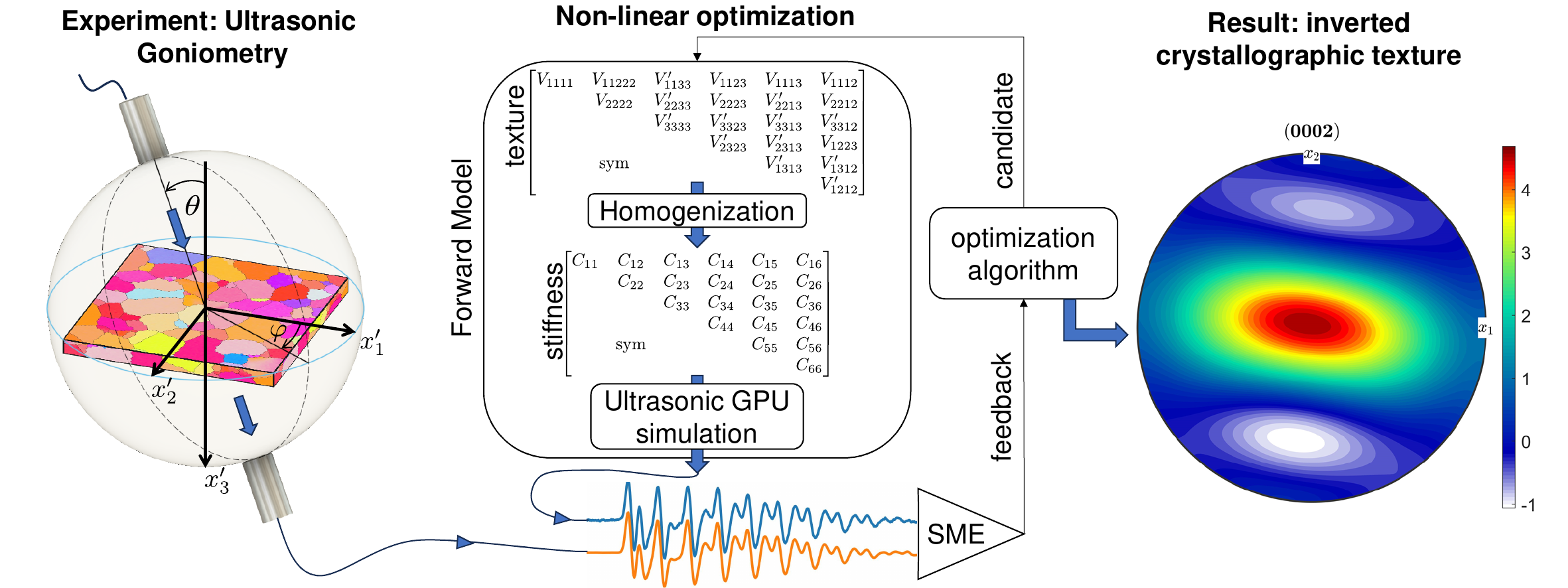}
\vfill
\end{graphicalabstract}

%%\item Research highlight 1
%%\item Research highlight 2


\end{frontmatter}

\section{notation}
Scalars are denoted by italic characters, e.g. $x,\ W$; first order tensors are denote by bold lower case letters, e.g. $\boldsymbol{x}$, $\boldsymbol{y}$; second order tensors are denoted by bold uppercase letters, e.g. $\boldsymbol{A}$, $\boldsymbol{B}$, except for the strain tensor $\epsilon$; and fourth order tensors are denoted by blackboard bold uppercase letters, e.g. $\mathbb{A}$, $\mathbb{B}$.  At the same time $r$-th order tensors are denoted as $\mathbb{A}_{\langle r \rangle}$. Linear mapping between a fourth and a second order tensor is denoted as $\boldsymbol{Ax}$, and the scalar product is denoted as $\mathbb{A}_{\langle r \rangle} \cdot \mathbb{B}_{\langle r \rangle}$. Single crystal properties are denoted as $\hat{\mathbb{C}}$, while effective properties are denoted as $\bar{\mathbb{C}}$. The local stiffness $\mathbb{C}(\boldsymbol{x})$, is abbreviated as $\mathbb{C}$. Angle brackets represents both the volumetric $\langle \mathbb{C} \rangle$ and the orientation $\langle f,\hat{\mathbb{C}} \rangle$ averages.
%}Apertura
\section{Introduction}
Crystallographic texture, defined as the orientation probability density function of crystals within a polycrystalline aggregate, is a critical parameter for the design, fabrication and diagnosis of engineering materials. It partially determines fundamental material properties, such as stiffness, strength, plasticity, magnetic susceptibility, chemical reactivity, and thermal conductivity \cite{Bunge1982}; therefore, during fabrication, it impacts formability, drawablity, weldability, among others \cite{Sowerby1975}; and later, during service life, it  governs degradation mechanisms such as irradiation growth, nodular corrosion, stress corrosion cracking, and hydride orientation precipitation \cite{LingaMurty2006}. 

Ultrasonic techniques for texture characterization have been extensively investigated over the past three decades due to their low cost, high penetration depth, and nondestructive nature. Unlike laboratory techniques, such as X-ray diffraction (XRD) and Electron Back-scattered Diffraction (EBSD), ultrasonic waves travel through the material providing a depth-averaged texture. This capability makes ultrasonic methods comparable with techniques only available at large scientific facilities such as neutron diffraction (ND) and Synchrotron X-ray diffraction (SXRD), while offering significantly greater accessibility and lower operational cost. Moreover, ultrasonic measurements are rapid, require minimal sample preparation, and can be implemented in industrial environments \cite{Lan2018}.

 The foundation of the method relies on the mechanical anisotropy of single crystals which, when combined with the orientation distribution of crystals, produce an unique direction dependent stiffness in the polycritalline sample. The spatial behavior of stress waves reflects this anisotropy, enabling inference of the effective stiffness tensor and, consequently, the crystallographic texture. Early studies focused on rolled sheet materials, employing Electro Magnetic Acoustic Transducers (EMAT) to exite (quasi-non-dispersive) plate wave modes(Lamb and shear horizontal) along multiple in-plane directions. The texture coefficients are inverted from the measured phase velocities by using explicit relations derived from the Voigt or Reuss bounds. Nevertheless, this relations restrict sample symmetry, and the propagation of plate waves imposes a maximum sample thickness.   
 
Lan et al. \cite{Lan2015hex, Lan2015cub, Lan2018}  proposed a new framework in which single crystal and aggregate velocities, treated as functions on the sphere, are expanded in spherical harmonics, where they are directly related by the texture coefficients via a convolution operation.
Although this formulation removes symmetry restrictions, it relies on a bulk-wave approximation to determine direction-dependent velocities, thereby requiring a minimum specimen thickness.
 
More recently, Rossin et al. \cite{Rossin2021} proposed an inversion strategy based on resonance ultrasonic spectroscopy (RUS), where texture coefficients are determined by minimizing the discrepancy between measured and predicted resonance frequencies. Their implementation incorporates second-order Hashin-Shtrikman bounds, following Lobos Fernández et al. \cite{LobosFernndez2018}, which improve upon classical Voigt and Reuss approximations by providing tighter bounds on effective elastic properties. Nevertheless, similar geometrical limitations remain because RUS requires parallelepiped specimens which cannot always be extracted from the material of interest, and which demand a significant effort during sample preparation. 

The present work aims to extend the applicability of ultrasonic texture determination by developing a methodology that imposes no specimen symmetry constraints while significantly reducing geometric restrictions. An immersion ultrasonic goniometry system is implemented to acquire transmitted fields over a range of controlled sample rotations. Rather than extracting phase velocities explicitly, a plane wave model is used to simulate the transmitted field, and texture coefficients are determined by minimizing the discrepancy between measured and modeled signals. The plane wave model, accounting for both fluid-solid interfaces, encompasses from plate to bulk waves and therefore it is applicable to samples over a wide range of thicknesses. The Hashin-Shtrikman theory is used to relate stiffness to microstructure, and vice-versa, following the recent advancements in the literature. The proposed approach is also favorable for elasticity determination because the stiffness tensor is a by-product of the regression. GPU implementation reduces computation time so that each inversion only takes a few minutes.

The method is validated on materials with cubic and hexagonal crystal symmetry for specimen thicknesses ranging from 0.5 to 5 mm. The inferred texture coefficients are compared with results obtained from XRD and ND measurements, demonstrating good agreement across a range of anisotropy levels, including weakly textured samples.

\section{Theoretical Basis}
% \textit{Early attempts at recovering texture from ultrasonic measurements were based on finding explicit relations between texture coefficients and elastic constants. Sayers defined the stiffness-texture relations for cubic \cite{Sayers1982} and hexagonal \cite{Sayers1986} materials by using the (first order) Voigt bound; Afterwards, the  stiffness-texture relations but using Voigt-Reuss-Hill average were implemented for cubic \cite{Hirao1987} and for hexagonal \cite{Li1990} materials. The convolution framework developed by Lan et al., while not directly related to the homogenization theories, was show to give close results to that obtained by the Voigt bound \cite{Lan2015hex}. The work of Rossin \cite{Rossin2021} used the tighter HS effective approximation as proposed by \cite{LobosFernndez2018} for RUS experiments in cubic materials. This works uses the same effective approximation for both cubic and hexagonal materials in traveling wave experiments.  
% }

% Sections \ref{sec:texture} and \ref{sec:homogeneization} follow the symbolic tensor notation and conventions used in \cite{LobosFernndez2018}.
\subsection{Elastic Homogenization}\label{sec:homogeneization}
\subsubsection{Zero, first and second order bounds}
Elastic homogenization consists on approximating an heterogeneous solid as a linear elastic homogeneous continuum. Based on the principles of minimum energy  and complementary energy, the effective material behavior $\Bar{\mathbb{C}}$ is delimited by the $n$th-order bounds $\mathbb{C}^{n}$, associated to $n$-point microstructure information as
\begin{equation}\label{eq:HS2}
\Bar{\epsilon}\cdot(\mathbb{C}^{n-}\Bar{\epsilon})\le \Bar{\epsilon}\cdot(\Bar{\mathbb{C}}\Bar{\epsilon})
\le\Bar{\epsilon}\cdot(\mathbb{C}^{n+}\Bar{\epsilon}),
\end{equation}
where $\Bar{\epsilon}$ is the effective strain. 
Zeroth-order bounds are independent of microstructure statistics and thus provide an isotropic limit regardless of texture \cite{Lobos2016}. First-order bounds are related to one-point microstructure statistics, and are usually referred to as Voigt bound \cite{Voigt1966} and Reuss \cite{Reuss1929} bounds, as
\begin{equation}\label{eq:HS5}
    \mathbb{C}^{1+}=\mathbb{C}^{\text{Voigt}}=\langle \mathbb{C} \rangle,\quad \mathbb{C}^{1-}=\mathbb{C}^{\text{Reuss}}=\langle \mathbb{C}^{-1} \rangle^{-1}.
\end{equation}
These are computed as the volumetric average of the stiffness $\langle \mathbb{C} \rangle$ and the inverse of the volumetric average of the compliance tensor $\langle \mathbb{C}^{-1} \rangle^{-1}$. Additionally, Hill \cite{Hill1963} proposed an average of the two in order to achieve a better estimation of the effective stiffness. 
Variational principles involving the polarization tensor have been introduced by Hashin and Shtrikman (HS) \cite{Hashin1962}, leading to higher order bounds. The HS expression,
\begin{equation}\label{eq:HS7}
    \mathbb{C}^{HS}=\mathbb{C}_0-\mathbb{P}_0^{-1}+\langle \mathbb{L} \rangle ^{-1}, \text{where} \ \mathbb{L}=[\mathbb{C}- \mathbb{C}_0 + \mathbb{P}_0^{-1}]^{-1},   
\end{equation}
depends on the comparison stiffness $\mathbb{C}_0$, and the polarization tensor $\mathbb{P}_0$.
Here, the assumption of isotropic two-point statistics (equiaxed crystals) leads to the a closed form definition of the polarization tensor $\mathbb{P}_0$. In consequence, the second-order HS bound for isotropic two-points statistics can be parametrized in terms of one-point statistics (volumetric fractions) of the microstructure. 
Using zeroth-order bounds as comparison stiffness $\mathbb{C}_0$ the second-order Hashin-Shtrikman bounds become
\begin{equation}\label{eq:HS10}
    \mathbb{C}^{2-}=\mathbb{C}^{HS}|_{\mathbb{C}_0=\mathbb{C}^{0-}},\quad  \mathbb{C}^{2+}=\mathbb{C}^{HS}|_{\mathbb{C}_0=\mathbb{C}^{0+}}.
\end{equation}
The effective stiffness, located within the HS bounds, can be can be approximated in two steps as proposed in \cite{LobosFernndez2018}. First, the isotropic self-consistent solution is computed from Eq. \ref{eq:HS7} but complying with the vanishing of texture $f(\boldsymbol{Q})=1$ (isotropic assumption). The resulting isotropic tensor $\mathbb{C}_0^{iso}$ is then re-introduced as the comparison stiffness in \ref{eq:HS7}, and the microstructure statistics are included to compute the effective stiffness approximation as
\begin{equation}\label{eq:HS12}
 \mathbb{C}^{HS}_{eff}= \mathbb{C}^{HS}|_{\mathbb{C}_0=\mathbb{C}_0^{iso}}.
\end{equation}
This tensor is analogous to the Hill approximation located between the Voigt and Reuss bounds, but this time obtained for the second-order HS bounds.

\subsection{Tensorial representation of texture}\label{sec:texture}
The orientation probability density function $f(\boldsymbol{Q})$ describes the volume fraction of crystals oriented by the rotation $\boldsymbol{Q} \in SO(3)$. It is commonly represented as a Fourier expansion in terms of Wigner D-functions \cite{Bunge1982}.
An alternative representation in terms of  $\alpha$-th order tensorial coefficients $\mathbb{V}_{\langle \alpha \rangle \beta}$ and orthogonal functions $\mathbb{F}_{\langle \alpha \rangle \beta}$, motivated by the fact that tensor transformations are well defined \cite{LobosFernndez2018}, can be expressed as
\begin{equation}\label{eq:tex2}
    f(\boldsymbol{Q})=\sum_{\alpha=0}^\infty \sum_{\beta=1}^{n_{\alpha}}(1+2\alpha)\mathbb{V}_{\langle \alpha \rangle \beta} \cdot \mathbb{F}_{\langle \alpha \rangle \beta} (\boldsymbol{Q}),
\end{equation}
where $n_\alpha=1+2\alpha$, unless crystal symmetries are applied. Traditional and tensorial coefficients can be translated between each other following \cite{LobosFernndez2018}.

\subsection{Harmonic decomposition}
The stiffness tensor can be decomposed into a direct sum of  subspaces of harmonic tensors through the harmonic decomposition (hd) \cite{Forte1996}. This decomposition has been extended for the case of polycritalline solids \cite{LobosFernndez2018}, by replacing the volumetric average $\langle  {\mathbb{C}}\rangle$ with the orientation average $\langle f, \hat{\mathbb{C}}\rangle$ and explicitly using the tensorial texture coefficients, $\mathbb{V}_{\langle \alpha \rangle\beta}$ , as
\begin{equation}\label{7.eq:desc_harm}
\begin{split}
    \mathbb{C}^{\text{Voigt}}&=\langle f, \hat{\mathbb{C}}\rangle \\
    &=
    \text{hd}(\hat{h}_{I1}
        ,\hat{h}_{I2}
        ,\sum\limits_{\beta=1}^{n_2}\hat{h}_{21 \beta} \mathbb{V}_{\langle 2 \rangle\beta}
        ,\sum\limits_{\beta=1}^{n_2}\hat{h}_{22 \beta} \mathbb{V}_{\langle 2 \rangle\beta}
        ,\sum\limits_{\beta=1}^{n_4}\hat{h}_{41 \beta} \mathbb{V}_{\langle 4 \rangle\beta}
        ),
        \end{split}
\end{equation}
where the $\hat{h}$ scalars depend on the single crystal stiffness $\hat{\mathbb{C}}$. This decomposition is exact, and shows that the homogenized elasticity depends only on coefficients of even order (2 and 4), as previously noted using the expansion on Wigner D-functions \cite{Bunge1965}. Moreover, the crystal symmetry is extended to the texture coefficients, which, for the hexagonal case ($n_2=n_4=1$), reduces Eq \ref{7.eq:desc_harm} to 
\begin{equation}\label{7.eq:voigt}
    \langle f,\hat{\mathbb{C}}^{hex}\rangle= 
    \text{hd}(\hat{h}_{I1} 
        ,\hat{h}_{I2}
        ,\hat{h}_{211} \mathbb{V}_{\langle 2 \rangle1}
        ,\hat{h}_{221} \mathbb{V}_{\langle 2 \rangle1}
        ,\hat{h}_{411} \mathbb{V}_{\langle 4 \rangle1}
        ).\\
\end{equation}
Note that the $\beta$ summations disappear, making Eq. \ref{7.eq:voigt} a one to one relation between stiffness and texture. This means that, provided access to the Voigt average, one can uniquely determine the texture coefficients involved. In practice, ultrasonic experiments access the effective stiffness $\bar{\mathbb{C}}$, which is closer to the the HS approximation, than the bounds ($|\bar{\mathbb{C}}-\mathbb{C}^{HS}_{app}|\leq|\bar{\mathbb{C}}-\mathbb{C}^{2\pm}| \leq |\bar{\mathbb{C}}- \mathbb{C}^{1\pm}|$). Therefore, Eq. \ref{eq:HS7} can be expressed, for an hexagonal crystal symmetry, as
\begin{equation}
    \begin{gathered}
         \mathbb{C}^{HS}_{app}=\mathbb{C}_0^{iso}-\mathbb{P}_0^{-1}+  \langle f,\hat{\mathbb{L}}\rangle^{-1},\\
    \text{with} \quad
     \langle f,\hat{\mathbb{L}}\rangle=\text{hd}(\hat{h}_{I1}
        ,\hat{h}_{I2}
        ,\hat{h}_{211} \mathbb{V}_{\langle 2 \rangle1}
        ,\hat{h}_{221} \mathbb{V}_{\langle 2 \rangle1}
        ,\hat{h}_{411} \mathbb{V}_{\langle 4 \rangle1}
        ),
\end{gathered}
\end{equation}
where the $\hat{h}$ scalars now depend on $\hat{\mathbb{L}}$. This relation remains bijective and only requires one orientation average. This represents an advantage over the VRH average which requires two. The analysis above also applies for materials with cubic crystal symmetry ($n_2=0$ and $n_4=1$), whose averages only depends on $ \mathbb{V}_{\langle 4 \rangle1}$. Therefore, texture for materials with lower crystal symmetry cannot be inverted by elastic methods.

\subsubsection{Independent texture coefficients}
The elastic behavior for cubic materials with triclinic sample symmetry is completely described by the fourth order texture coefficient  $ \mathbb{V}_{\langle 4 \rangle1}$ with 9 independent components \cite{Bhlke2014}, represented in normalized Voigt notation as
\begin{equation}\label{eq:HS13}
   \boldsymbol{V}_ 4=
    \begin{bmatrix}
    V_{1111}&V_{11222}&V_{1133}'&V_{1123}&V_{1113}&V_{1112}\\
    \,    &V_{2222}&V_{2233}'&V_{2223}&V_{2213}'&V_{2212}\\
    \,    &\,    &V_{3333}'&V_{3323}'&V_{3313}'&V_{3312}'\\
    \,      &\,   &\,    &V_{2323}'&V_{2313}'&V_{1223}\\
   \,     &\text{sym}   &\,    &\,   &V_{1313}'&V_{1312}'\\
    \,    &\,    &\,    &\,    &\,    &V_{1212}'
    \end{bmatrix}
\end{equation}

Here, linearly dependent components are denoted by an apostrophe. The elastic behavior for hexagonal materials with triclinic sample symmetry, also depends on $ \mathbb{V}_{\langle 4 \rangle1}$, and on a second order texture coefficient  $ \mathbb{V}_{\langle 2 \rangle1}$ with 5 independent components, denoted here as

\begin{equation}\label{eq:HS14}
    \boldsymbol{V}_ 2=
    \begin{bmatrix}
    V_{11}&V_{12}&V_{13}\\
    \,    &V_{22}&V_{23}\\
    \text{sym}&\   &V_{33}'
    \end{bmatrix}.
\end{equation}

Moreover, sample symmetry reduces the number of independent components in $\boldsymbol{V}_ 4$ and $\boldsymbol{V}_ 2$.

\subsubsection{Bound visualization}
\begin{figure}[!ht]
    \centering
    \includegraphics[width=0.95\linewidth]{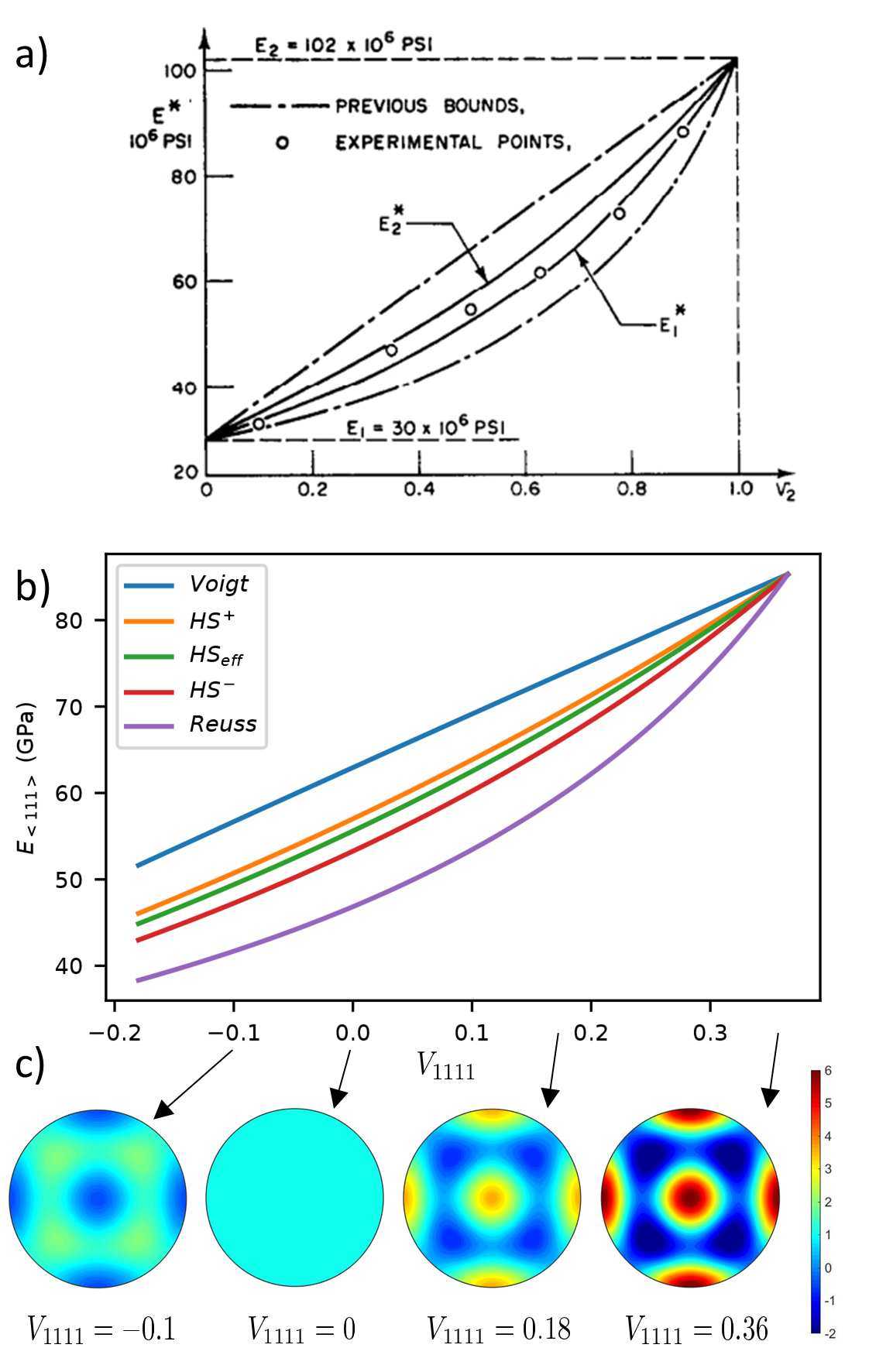}
    \caption{Elastic Homogenization. a) Bounds for biphaseic materials \cite{Hashin1963}. b) Bounds for the Young modulus in the 111 direction for a cubic-cubic material. c) $\langle 100 \rangle$ polar plots for different values of $V_{1111}$.}
    \label{fig:bounds}
\end{figure}

The bounds are classically represented for biphasic materials, for example in \cite{Hashin1963}, included in Fig. \ref{fig:bounds}.a.
A material with cubic crystal and sample symmetries (cubic-cubic) depends only on a single fourth order texture component $V_{1111}$, therefore it is useful for visualizing elastic bounds for policrystalline materials. First and second order bounds together with the HS approximation were computed for a stainless steel whose single crystal elastic constants are obtained from \cite{Ledbetter1985}. The homogenized Young modulus for the 111 direction ($\varphi=45^{\circ}$, $\theta=54.74^{\circ}$) $E_{<111>}$ as a function of $V_{1111}$ is shown in Fig.\ref{fig:bounds}.b. 
 It can be seen that at $V_{1111}=0.36$, the bounds concur to the same point, because the material can be tough of as an homogeneous single crystal. As $V_{1111}$ diminishes, the material becomes heterogeneous and the bounds bifurcate. The $<100>$ polar plots for some points are included in Fig.\ref{fig:bounds}.c. A singular plot corresponds to the isotropic state ($V_{1111}=0$) which is a subset of the cubic symmetry group. 

\subsection{Ultrasonic triclinic plane wave model}\label{sec:plane-wave}
Most works regarding ultrasonic determination of crystallographic texture measure bulk or plate wave velocities which, through the Christoffel equation, are related to the stiffness tensor, which in turn is related to the texture coefficients via homogenization methods. The disadvantage of this approach is that bulk wave velocity cannot be determined in plate samples and plate wave velocity cannot be determined in thick samples. Moreover, velocity determination carries some challenges, such as mode determination or phase distortion compensation, which may hinder accuracy. The method proposed in this work consists on fitting the measured signal with a plane wave model which encompasses different propagation modes and a wide range of specimen thicknesses. Because of this, explicit velocity parametrization is avoided. The plane wave model is briefly discussed here, but the complete description can be found in \cite{cowes2}. 

Christoffel equation solves phase velocities $v$ for propagation direction $\boldsymbol{n}$ in linear elastic homogeneous generally anisotropic media. At a fluid-solid interface, due to refraction, it's not possible to know $n_j$ a priori. Conversely, if the wave impinges from the fluid at an angle $\theta$ from the interface normal, the wavevector component, $k_1=|\boldsymbol{k}^i|\sin(\theta)$ is common to both media according to the Snell's law, and the unknowns become the vertical wavevector components $k_3^q$. Here, $q$ refers to one of the six propagation modes: three traveling upwards and three traveling downwards; and $|\boldsymbol{k}^i|=\omega/v_f$, with $\omega$ being the angular frequency and $v_f$ is the velocity in the fluid. 
\begin{figure}
    \centering
    \includegraphics[width=0.8\linewidth]{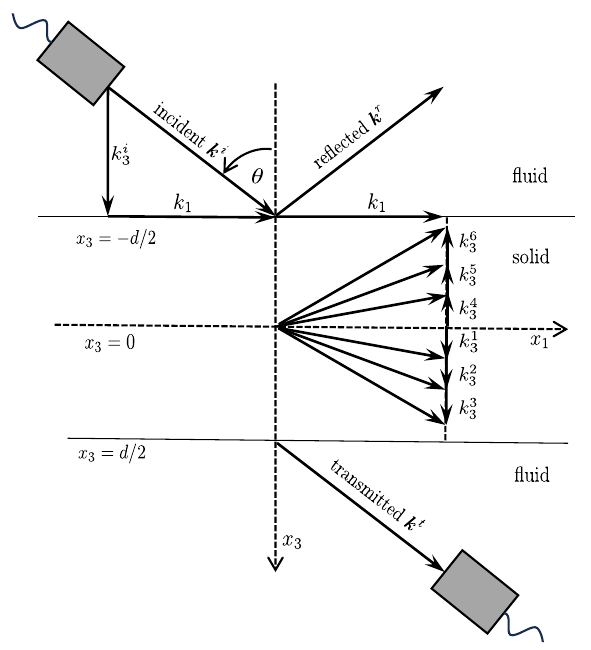}
    \caption{Partial waves and wavevectors in a fluid immersed triclinic plate.}
    \label{fig:scheme_transmission}
\end{figure}
Because of a coordinate transformation for each azimuthal angle, motion in the plane $(x_1,x_3)$ is independent of $x_2$ and the solutions for the displacement equations can be sought in the form 
\begin{equation}\label{eq:w7}
    \mathbf{u}=\mathbf{a} \exp{(i(k_1 x_1+k_3^q x_3-\omega t))}, \,\,\,\,  
\end{equation}
where $\mathbf{u}$ is displacement, $\mathbf{a}$ is wave polarization, and $t$ is time. Substituting eq. \ref{eq:w7} into the Navier's stress-displacement equation leads to the modified Christoffel equation, as
\begin{equation}\label{eq:w8}
\mathbf{K}(k_3^q)\mathbf{a}=0, 
\end{equation}
The determinant of $\mathbf{K}$ must vanish for Eq. \ref{eq:w8} to have non trivial solutions, leading to a polynomial whose roots are the wavevector components $k_3^q$, which are subsequently used to determine the amplitudes $\mathbf{a}$. For a fluid immersed solid plate, continuity of the normal stress ( $\bar{\sigma}_{33}=\sigma_{33}$ ) and displacement ( $\bar{u}_3=u_3$ ) at each interface is assumed, while shear stress must vanish ($\sigma_{13}=\sigma_{33}=0$) because the fluid does not support shear forces. Here, the bar indicates a quantity in the fluid and no bar indicates a quantity in the solid. These continuity conditions lead to an 8 by 8 system of linear equations which, upon solving, lead to the transmission coefficient $T$.

Finally, the temporal signal can be simulated by performing an inverse Fourier transform as
\begin{equation}\label{eq:w15}
s(\theta, \phi, t)=\int_{-\infty}^{\infty} T(\theta, \phi, \omega)\Gamma(\omega)\exp(i\Phi)\exp(i\omega t)d\omega,
\end{equation}
 where $\Gamma$ is the transfer function of the system and the term $\exp(i\Phi)$ accounts for the phase difference which occurs when introducing the solid plate.
 
\begin{figure}
    \centering
    \includegraphics{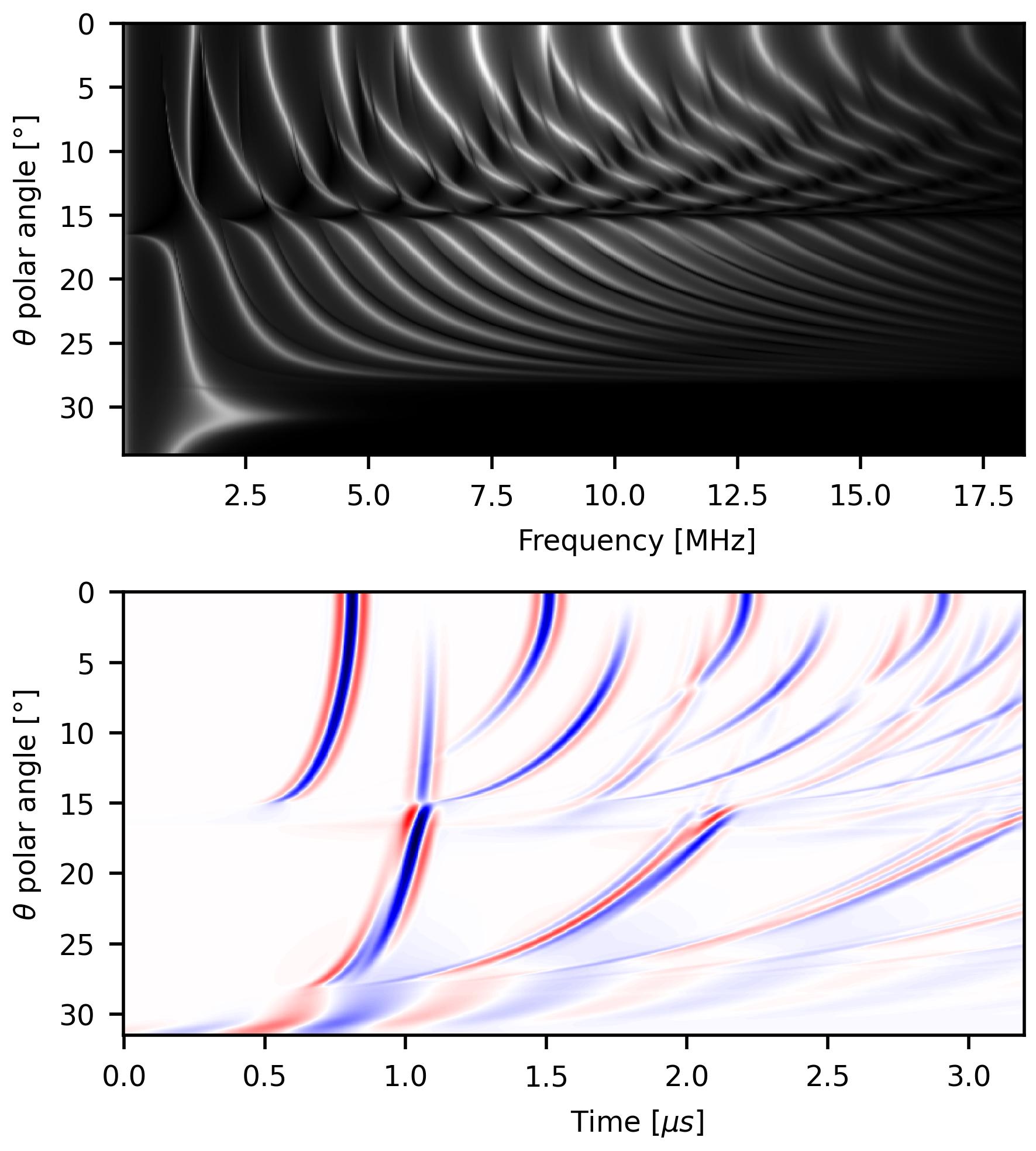}
    \caption{Plane wave simulation for a triclinic steel plate for a polar scan. Top: magnitude of the transmission coefficient. Bottom: transmitted signal}
    \label{fig:sim_t_f}
\end{figure}

A simulation for a triclinic 304 plate with a thickness of 2 mm is shown in Fig. \ref{fig:sim_t_f}. The transmission coefficient $|T|$ and the transmitted signal $s$ are shown for a single polar scan at fixed azimuthal angle.

\section{Experimental method}
\subsection{Ultrasonic goniometry}
\begin{figure}
    \centering
    \includegraphics[width=0.85\linewidth]{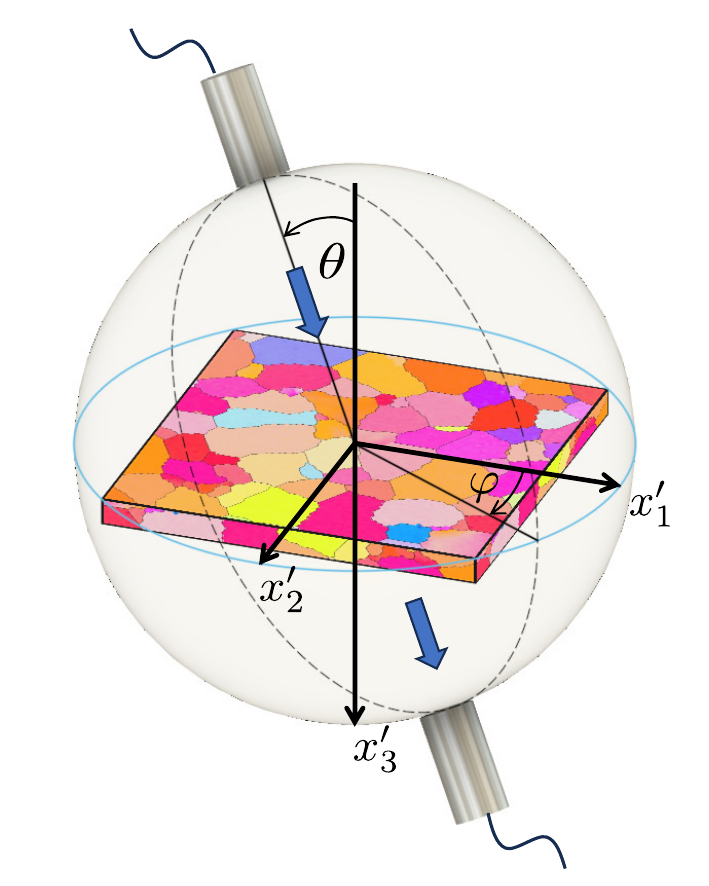}
    \caption{Ultrasonic goniometry diagram. An ultrasonic field impinges a polycrystalline sample while varying angles of incidence.}
    \label{fig:goniometry}
\end{figure}
The ultrasonic goniometry experiment implemented here consist of rotating a sample located between two transducers under immersion as shown in Fig. \ref{fig:goniometry}. Rotation is performed by continuous polar scans at 12 different azimuthal angles. This number is founded on the spherical harmonic sampling theorem, used in \cite{Lan2018}, which determines the minimum sampling step to correctly acquire spherical data. 

Plane wave assumption is required for the application of the model presented in section \ref{sec:plane-wave} . Ultrasonic goniometry experiments usually generate a lateral field displacement due to leaky guided wave propagation, which finite transducers fail to acquire entirely, voiding plane wave assumption. A finite beam model can simulate such a setup but commonly require the double integration for all the wavevectors of interest and as such it is computationally prohibitive for the numerous evaluations during the inversion.  Dedicated PVDF transducers have been developed for the use in ultrasonic goniometry experiments by an optimization process which found the transducer shape and position which better resemble plane wave propagation. The fabricated transducers were successfully validated by experiments in anisotropic media \cite{Cowes2024}. 

Media homogeneity assumption is also required for both the plane wave model and the elastic homogenization. This can be voided at high frequencies where grain size becomes comparable to wavelength making grain scattering control wave propagation. More complex equations can account for grain shape and size while simultaneously considering elasticity \cite{Sha2018}, but become computationally expensive as well. 

The Gaussian frequency response of the transducer naturally weighs the frequencies of interest diminishing the influence of grain scattering at high frequencies and beam diffraction at low frequencies. In this work, a central frequency of 10 MHz is adequate for the grain size of the worked metals tested.

% HS homogenization, to be described by texture coefficients only, requires equiaxed crystals, and the absence of residual stress, which is not accounted for in the model.

\subsection{Optimization}\label{sec:inversion}
Because the equations that relate crystallographic textures to transmitted ultrasonic fields are not analytically invertible a fitting approach is needed by which the inputs to a forward model are iterated looking to minimize the error between the experimental data and the model. Gradient based optimizations are used when the objective function is differentiable, and the derivatives of the function ares used to guide the steps in search for the minima. Otherwise, metaheuristics alternatives propose a set of rules, which, while do not guarantee the convergence to a global minimum, are independent of the function derivatives. Among these, the Pattern Search algorithm was selected due to its effectiveness for multi-variable global optimization \cite{Hooke1961}, using the Python Pymoo library \cite{pymoo} implementation.

\begin{figure}
     \centering
     \includegraphics{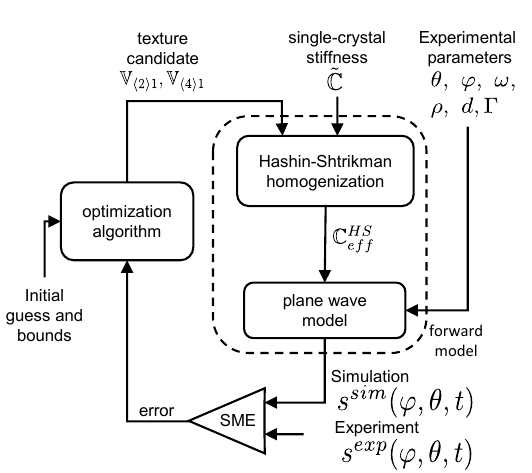}
     \caption{Inversion of the experimental data. Texture coefficients are varied until the simulation error is minimized.}
     \label{fig:opti-scheme}
 \end{figure}

A forward model is used to simulate the transmitted field from texture coefficients, and the error between simulation and measurement is used by the algorithm to adjust the texture coefficients for the subsequent population. A scheme of the process is show in Fig. \ref{fig:opti-scheme}. The result of the inversion process is the set of texture coefficients $\boldsymbol{V}_{2},\boldsymbol{V}_{4}$ which best fit the experimental data. The polycrystalline stiffness tensor is also obtained as a by-product.

A two step approach, first optimizing for stiffness on the signals, and then optimizing for texture on the stiffness, could be implemented , as used in \cite{Lan2018_rus}. Nevertheless, a combined approach is preferred because the search space is constrained by the realizable microstructures, improving stability and convergence during optimization \cite{Rossin2021}.

The objective function is defined for a candidate texture ($\boldsymbol{V}_{2},\boldsymbol{V}_{4}$) as the squared mean error (SME)
\begin{equation}
    F(\boldsymbol{V}_{2},\boldsymbol{V}_{4})=\frac{1}{N}\sum_{i=0}^N(s_i^{sim}(\boldsymbol{V}_{2},\boldsymbol{V}_{4})-s_i^{exp})^2,
\end{equation}
where, $s^{sim}$ and $s^{exp}$ are the simulated and experimental signals respectively, the subindex $i$ goes over the variables $\varphi,\theta,t$, and $N$ is the total sampled points.

\subsubsection{Forward Model computation}
The forward model consists of the computation of a single individual which starts from a set of texture coefficients and ends with a set of transmitted fields. It can be separated in two parts, the stiffness homogenization and the ultrasonic simulation. 

The stiffness homogenization is implemented by using a modified version of the Rossin repository \cite{githubGitHubJrossintextureRUS}, which is a Python translation for the Mathematica code written by Lobos Fernández et. al.\cite{LobosFernndez2018}. Two modifications were introduced for the present work: First, the hexagonal crystal system was added. Secondly, as the HS effective approximation has parameters which only depend on single crystal elastic constants, these are calculated only once and stored. Then, for each new individual the set of texture coefficients is used to calculate the orientation average in \ref{eq:HS12}. This saves considerable time by avoiding redundant operations. The homogenization operation for each individual takes only $0.2\ ms$ in a 3.5 GHz 6 core CPU.

The second step starts with HS approximation tensor to simulate the transmitted signal. As the code consists of a great number of repeated operations, it is highly parallelizable which means a that great time savings can be attained by implementing it graphical processing units (GPUs). In the current implementation the ultrasonic simulation for an individual with a dimension of 12 $\phi$ by 200 $\theta$ by 256 $t$ takes approximately $50 ms$ in a commercial GPU. The optimization of the forward model is crucial to limit inversion time, due to the high number of model evaluations during fitting.

\subsubsection{Initial guess and bounds}
All texture coefficients are constrained by their Frobenius norm $||\mathbb{V}_{\langle\alpha\rangle \beta}|| \in [0,1]$. This set can also be interpreted as a convex hull of all possible single crystal states.  Because of this, the convex hull determined for a material with crystal and sample symmetry is a subset of the Frobenius norm \cite{LobosFernndez2018}, and therefore constitutes a smaller search domain.
Following this approach, the texture coefficients for 5000 randomly orientation single crystals were calculated for cubic and hexagonal crystals. The minimum and maximum values were stored for their posterior use as texture bounds during inversion, see section \ref{sec:inversion}. These values are included in the appendix(SECTION).
% Alternativa
% Following this approach, the texture coefficients for 5000 single crystals randomly orientation where calculated for cubic and hexagonal crystals. Each texture coefficient is only valid if it can be expressed as a linear combination of the convex hull points. This is used during inversion to constrain texture coefficients by checking the feasibility of the linear combination as a linear programming problem.

\subsection{Samples}
Several samples were chosen to test for different crystal symmetry, composition and thickness. All materials have a single phase microstructure, and have two parallel faces as a result of rolling. No sample preparation was done other than cutting to fit in the ultrasonic goniometer. The surfaces were not polished for the ultrasonic measurements. The sample details are included in table \ref{tab:sample_details}. Results are referred to as the measurement technique followed by the sample number, ie. $US-304_b$, refers to the results obtained by ultrasound for the $304_b$ sample.

% \usepackage{booktabs}
% \usepackage{multirow}
% \usepackage{graphicx}

\begin{table*}[!ht]
\caption{Sample details}
\label{tab:sample_details}
\centering
\resizebox{1.9\columnwidth}{!}{%
\begin{tabular}{@{}cccccccc@{}}
\toprule
\multirow{2}{*}{Sample} & \multirow{2}{*}{Material} & \multirow{2}{*}{Micro-simmetry} &\multirow{2}{2cm}{\centering Thickness (mm)} &  \multirow{2}{2cm}{\centering Density   $(\text{g}/\text{cm}^3)$} & \multirow{2}{2cm}{\centering Comparison technique} & \multirow{2}{2cm}{\centering AF (single cristal)} & \multirow{2}{2.5cm}{\centering Single crystal elastic   constants} \\
 &  &  &  &  &  &  &  \\ \midrule
$\textbf{Zry}$ & Zircaloy 4 &  hexagonal &5.00 & 6.50 & ND & 1.61 & \cite{Fisher1964} \\
$\textbf{SS304}$ & AISI 304 steel& cubic & 4.94  & 7.93 & ND & 3.18 &  \cite{Ledbetter1985}\\
$\textbf{Zn}$ & VM Zinc & hexagonal& 0.82  & 7.18 & XRD & 18.91 & \cite{Alers1958} \\
$\textbf{Si}{<}111{>}$ & Silicon ${<}111{>}$& cubic & 0.53  & 2.33 & XRD & 9.4 &  \cite{Hall1967}\\ \bottomrule
\end{tabular}%
}
\end{table*}

\subsection{Comparison techniques}
The ultrasonic measurements were compared against conventional techniques. Thin ($\approx 1 mm$) samples were measured by XRD and thick samples ($\approx 5 mm$) were measured by ND at the Australian Centre for Neutron Scattering. This decision was motivated by the fact that the neutron beam size requires a minimum thickness, which thin samples do not comply. DRX penetration depth is small, in the order of the micron, and does not provide trough thickness information.
DRX and ND pole figures were inserted into MTEX \cite{Bachmann2010} to compute the ODF, from which the traditional coefficients up to the fourth order were extracted. Truncated pole figures were then constructed from this coefficients.

The tensorial texture coefficients $\mathbb{V}_{\langle \alpha \rangle \beta}$ obtained by the ultrasonic inversion were translated to the  conventional coefficients, which then were used in MTEX to generate the same pole figures as above.
The confidence interval is 

\section{Results}
\subsection{Field fitting}
\begin{figure}[ht]
    \centering
    \includegraphics[width=0.9\columnwidth]{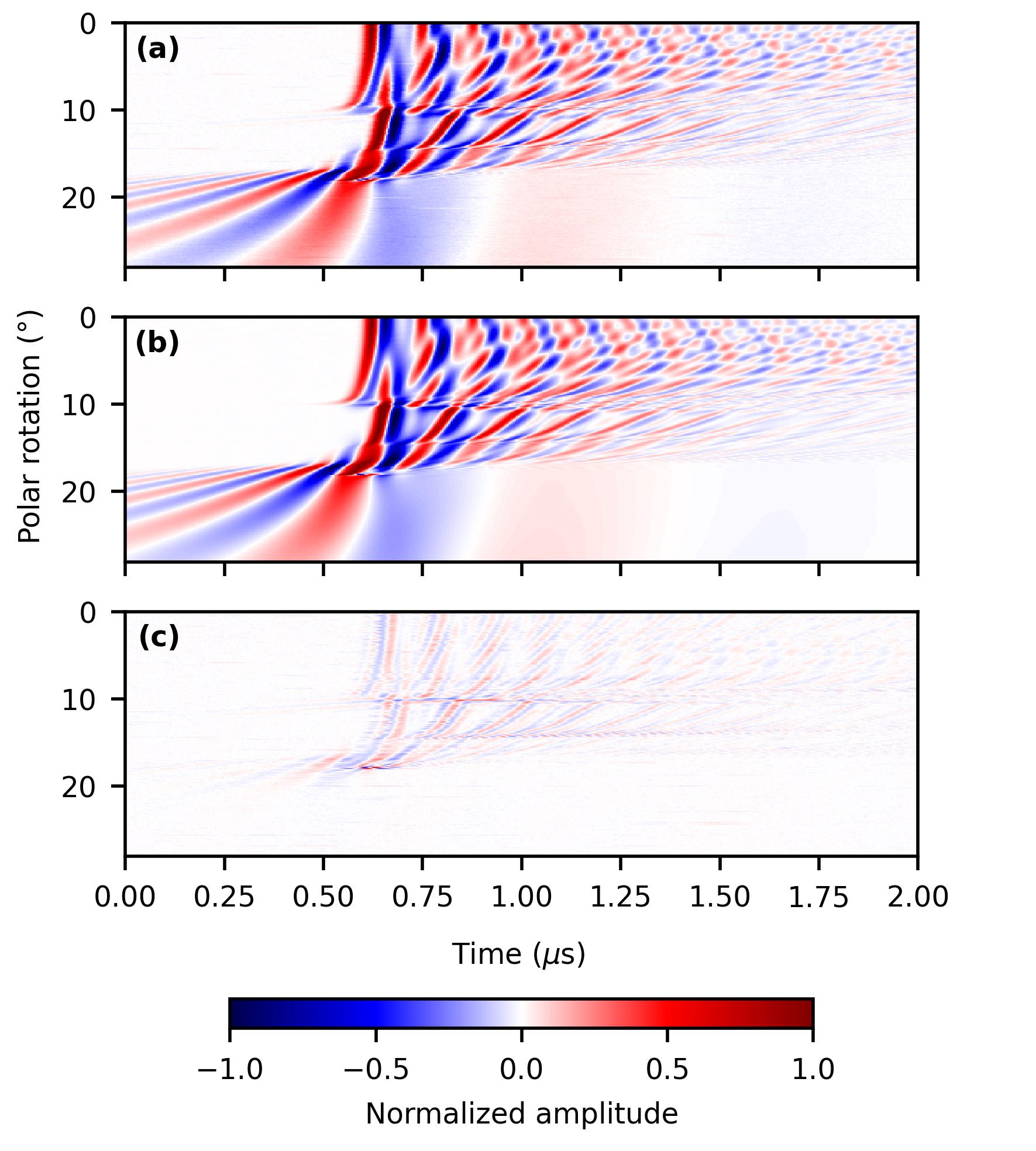}
    \caption{Comparison of experimental and fitted signals, for a single polar scan ( $\varphi=10^{\circ}$ ) in the $\textbf{Si$<$111$>$}$ sample. (a) experimental signal. (b) fitted signal. (c) subtraction of (a) and (b).}
    \label{fig:comp_scan}
\end{figure}

\begin{figure}[ht]
    \centering
    \includegraphics[width=0.9\columnwidth]{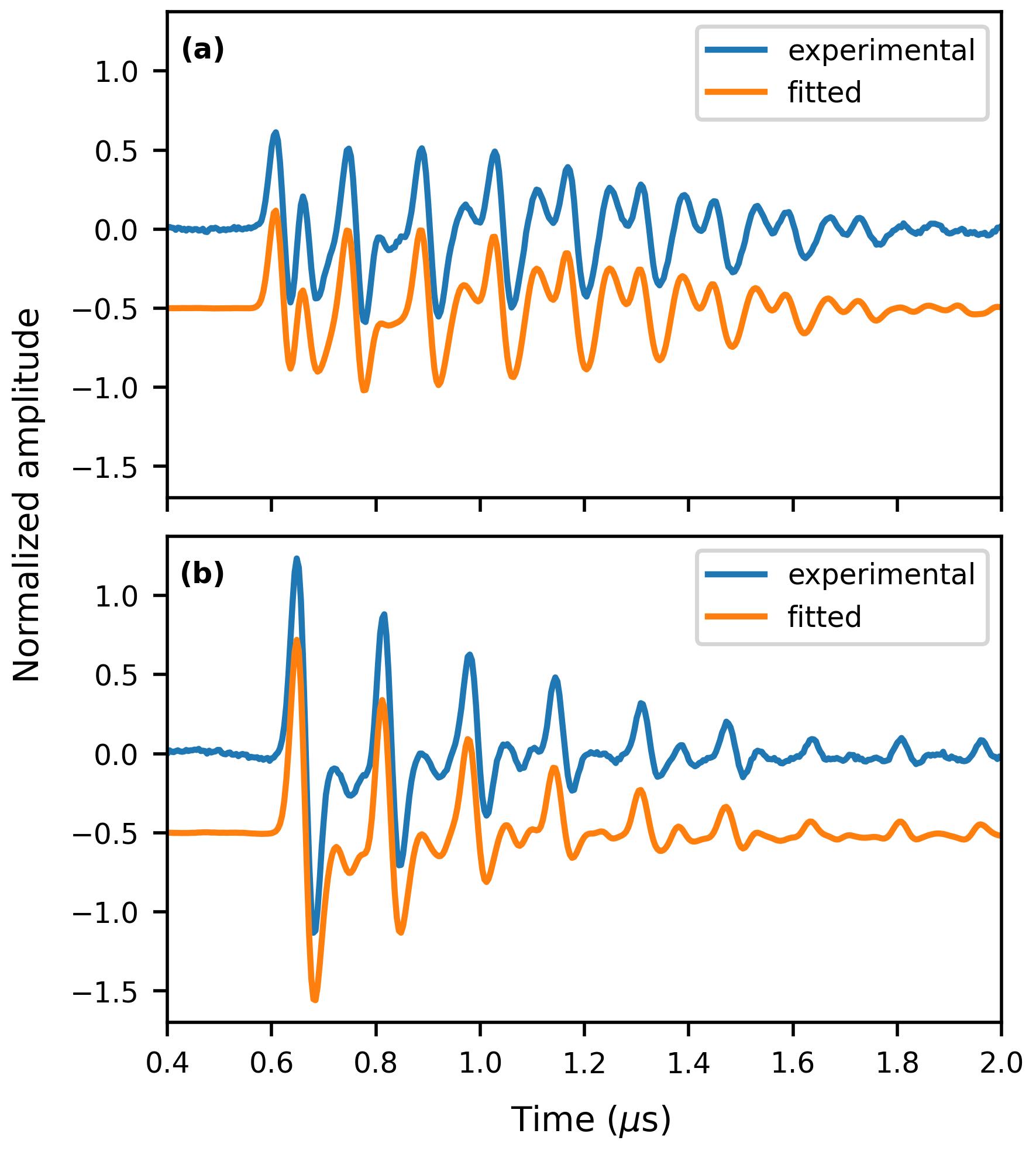}
    \caption{Comparison of experimental and fitted signals ( $\varphi=10^{\circ}$ ) for the $\textbf{Si$<$111$>$}$ sample. (a) $\theta=5.7^{\circ}$. (b) $\theta=11.3^{\circ}$. The fitted signals have an offset of -0.5 for better visualization.}
    \label{fig:comp_t}
\end{figure}

The optimization process minimizes the error between simulated and measured signals. The fitted field is compared to the experiment for the $430$ sample in Fig. \ref{fig:comp_scan}, where a single polar scan is displayed. Fig. \ref{fig:comp_t} shows the same comparison but only for two~$\theta$ polar angles. The similarity observed indicates that the model adequately describes the experiment.

\subsection{Pole figures}
\begin{figure*}[!ht] 
    \centering
    \includegraphics[width=1.5\columnwidth]{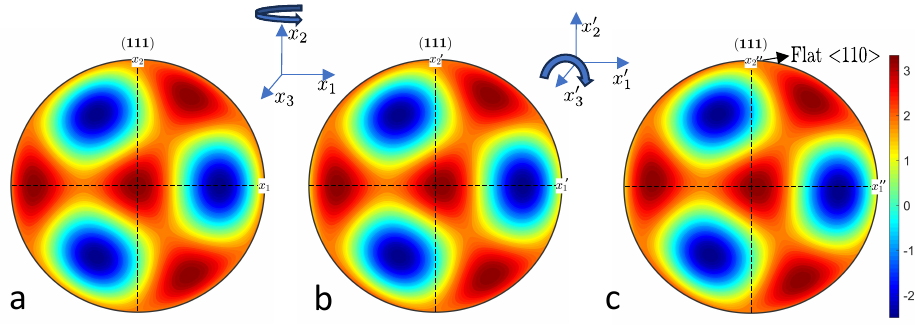}
    \caption{Misorientation of the $\mathbf{Si\langle 111\rangle}$ sample observed by ultrasound. Starting from the reference orientation (a), the misorientation can be described as a rotation of 4.2 $^{\circ}$ with respect to the $x_2$ direction (b) and a rotation of -3.2 $^{\circ}$ with respect to $x_3'$ (c).}
    \label{fig:si_mis}
\end{figure*}

The $\mathbf{Si\langle 111\rangle}$ sample was chosen as a reference since, being a single crystal, its texture components and stiffness tensor are accurately known. However, when performing the ultrasonic measurement, a deviation with respect to the expected orientation was found. Figure \ref{fig:si_mis} shows the (111) pole figure, where, starting from the reference orientation, the misorientation can be described as a rotation of 4.2 $^{\circ}$ with respect to the $x_2$ direction and a rotation of -3.2 $^{\circ}$ with respect to $x_3'$. These results were confirmed by XRD with an error smaller than 1$^\circ$ in each direction.

\begin{figure}[!ht] 
    \centering
    \includegraphics[width=.97\columnwidth]{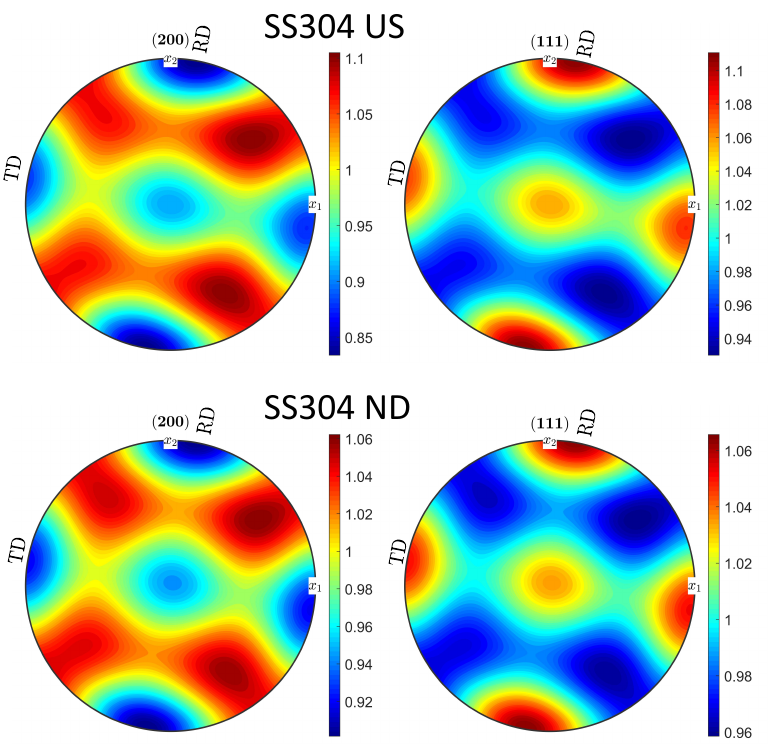}
    \caption{(200) and (111) pole figures for the $\textbf{SS304}$ sample obtained by the ultrasonic method (US) and neutron diffraction (ND). RD and TD represent the rolling and transverse directions, respectively, while $x_1$, $x_2$, and $x_3$ correspond to the measurement reference system.}
    \label{fig:pf304}
\end{figure}

\begin{figure}[!ht] 
    \centering
    \includegraphics[width=.97\columnwidth]{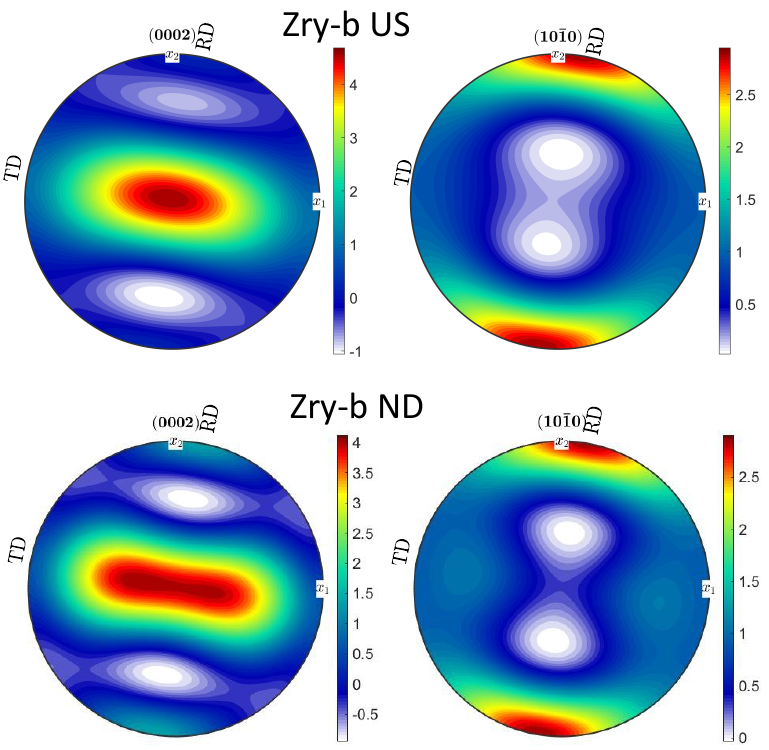}
    \caption{(0002) and $(10\bar{1}0)$ pole figures for the $\textbf{Zry}$ sample obtained by the ultrasonic method (US) and neutron diffraction (ND). RD and TD represent the rolling and transverse directions, respectively, while $x_1$, $x_2$, and $x_3$ correspond to the measurement reference system.}
    \label{fig:pfzryb}
    \end{figure}

\begin{figure}[!ht] 
    \centering
    \includegraphics[width=.97\columnwidth]{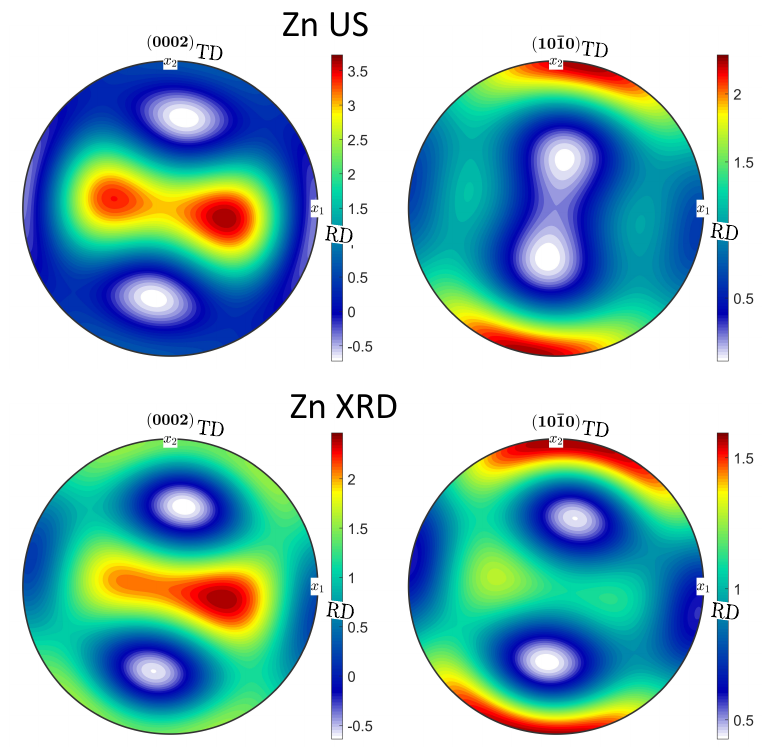}
    \caption{(0002) and $(10\bar{1}0)$ pole figures for the $\textbf{Zn}$ sample obtained by the ultrasonic method (US) and X-ray diffraction (XRD). RD and TD represent the rolling and transverse directions, respectively, while $x_1$, $x_2$, and $x_3$ correspond to the measurement reference system.}
    \label{fig:pfzn}
\end{figure}

The results obtained for the polycrystalline samples, $\textbf{SS304}$, $\textbf{Zry}$, and $\textbf{Zn}$, are shown in Figs. \ref{fig:pf304}, \ref{fig:pfzryb}, and \ref{fig:pfzn}, respectively. Once again, the alignment of the principal directions of the samples with the ultrasonic measurement reference system was intentionally avoided in order to test the flexibility of the method. Consequently, the comparison measurements were rotated to match the ultrasonic measurements. Nevertheless, the principal directions, RD and TD, are indicated in the figures.

The relative error between ultrasonic and diffraction based crystallographic texture was computed using the ``calcError" function in MTEX, and the results are included in Table \ref{tab:error_tex}.

\begin{table}[ht]
\centering
\caption{Relative error between ultrasonic and diffraction based crystallographic texture.}
\label{tab:error_tex}
\begin{tabular}{@{}clcc@{}}
\toprule
Sample                & \multicolumn{1}{c}{Error} & Comparison technique & Macroscopic AF \\ \midrule
$\textbf{Zry}$       & 0.23                      & ND                    &1\\
$\textbf{Zn}$          & 0.48                      & XRD                 & 2 \\
$\textbf{SS304}$       & 0.03                      & ND                 &  3 \\
$\textbf{Si{<}111{>}}$ & 0.04                     & XRD                 &  4\\ \bottomrule
\end{tabular}
\end{table}

\subsection{Kearns factors}
Anisotropic growth under fission fragment irradiation observed on Zirconium alloys motivated the development of a model for the calculation of properties influenced by texture \cite{kearns1965report}. This theory included the definition of the Kearns factors which quantify the volumetric fraction of basal poles directed towards each of the three principal directions of a specimen. Since the computation of these factors can be carried out by using only the texture coefficients of 2nd degree , the ultrasonic method is sufficient for their calculation \cite{Anderson1999}. Kearns factors were computed on MTEX from the ultrasonic measurements of the Zirconium based samples. The results are included in Table \ref{tab:kearns} for the rolling, tangential and normal directions. A better agreement is found for the $\textbf{Zry}$ sample than for the $\textbf{Zn}$ sample. This migth be explained by the fact that both ultrasonic and neutron methods are volumetric, while X-rays are shallow.

\begin{table}[!ht]
\centering
\caption{Kearns factors for the hexagonal samples.}
\label{tab:kearns}
\begin{tabular}{@{}ccccc@{}}
\toprule
 &  \multicolumn{2}{c}{$\textbf{Zry}$} & \multicolumn{2}{c}{$\textbf{Zn}$} \\ \midrule
Direction & US    & ND   & US   & DRX  \\ \midrule
Rolling       & -0.01 & 0.07 & 0.31 & 0.35 \\
Transverse        & 0.34  & 0.32 & 0.12 & 0.27 \\
Normal     & 0.66  & 0.61 & 0.57 & 0.38 \\ \bottomrule
\end{tabular}
\end{table}

\section{Discussion}
Very good agreement is found between ultrasonic and diffraction based techniques, proving that the developed framework for the ultrasonic determination of texture is successful. Samples with thickness raging from 0.23 to 5 mm were tested showing that bulk wave assumption is not required for this development. Absence of odd order coefficients can generate a "ghost" effect in tho ODF reconstruction but does not impact pole figures. On the other hand fourth order truncation can generate visual discrepancies compared to the full bandwidth coefficients\cite{Lan2018}, particularly for samples with sharp textures. For this cases the retrieved information still describes elasto-plastic  and thermomechanical behavior completely\cite{lobos2018homogenization}, as well as some magnetic properties \cite{Bunge1989}.
The differences in the samples may be attributed to residual stress, failing to comply with the equiaxed grain assumption or errors in the single crystal elastic constants which are taken from the literature. parallelism in thick samples and bowing in thin samples may also hinder with the accuracy of the method.

Waveform fitting shows that it is a viable alternative to velocity based techniques because it contemplates displacement amplitudes and non homogeneous waves so that fewer assumptions are made. The higher computational burden is counteracted by the efforts in making the forward model computationally efficient, particularly the GPU implementation of the ultrasonic model. The whole inversion for a single sample takes less than 7 minutes. Moreover the goniometry apparatus developed for this work permits the complete sampling of the transmitted field in under two minutes. 

The inversion performed by coupling both steps (Elastic Homogenization and Plane Wave Model) has the following advantages: First, the set of possible texture components are bounded by the single crystal symmetry. The homogenization strategy, by combining texture coefficients with the single crystal elastic constant in turn constrains the set of independent elastic constants. Therefore, the search space during inversion is greatly reduced which improves converge as noted by Rossin \cite{Rossin2021}. Moreover, it ensures that the fitted field is consistent with a textured aggregate of the material of interest.

\subsection{Limitations}
The elastic behavior for cubic materials with triclinic sample symmetry is completely described by a fourth order texture coefficient  $\boldsymbol{V}_{4}$ with 9 independent components \cite{LobosFernndez2018}. Because of this the information which can be retrieved by ultrasonic inversion is limited to this fourth order basis tensor. However, due to cubic crystal symmetry the first, second and third order coefficients are always null, so that the ultrasonic inversion completely describes texture coefficients up to the fourth-order.

On the other hand, the elastic behavior for hexagonal materials with triclinic sample symmetry is completely described by a second order texture coefficient  $\boldsymbol{V}_{2}$ with 5 independent components and  a fourth order texture coefficient  $\boldsymbol{V}_{4}$ with 9 independent components \cite{LobosFernndez2018}. This time, first and third order coefficients do exist, but cannot be retrieved by ultrasonic inversion due to the centrosymmetry of the elasticity. This is the same for diffraction experiments and is the origin of the ``ghost effect" which appears during ODF reconstruction \cite{Matthies1979}. Pole figures are independent of odd order coefficients as well.

\section{Conclusion}
A method for the determination of crystallographic texture by ultrasonic goniometry experiments was presented. The use of a wave fitting approach instead of velocity based techniques permitted the inversion of results for specimens of varying thicknesses, without making bulk or plate wave assumptions. Moreover, no sample symmetry was imposed. 

The application of HS elastic homogenization delivered a tight coupling between specimen texture and elastic stiffness, which aided inversion convergence and improved the obtained results. 

Good agreement was found for hexagonal and cubic samples when compared to X-ray and neutron diffraction methods. The GPU implementation of the triclinic plane wave model was critical in reducing the inversion time during the meta heuristic optimization procedure. With the implemented improvements, results can be obtained in less than 10 minutes accounting for measurement and inversion. The fact that almost none sample preparation is required, that the information is thickness averaged coupled to the short time taken by the experiment, makes the presented technique an attractive alternative to traditional methods.

% COSAS QUE FALTAN:

% agregar a las tablas de resultados optimizados los errores en el ajuste
% agregar los af macroscópicos

% \appendix
% Por último, como se describió en la sección \ref{sec:anomalous}, la ecuación de Christoffel, si incluye información de polarización, también es inyectiva. Sin embargo, como se mostró en la sección \ref{sec:unicidad}, la goniometría ultrasónica presenta incertidumbre para materiales con macro-simetría hexagonal alineada con el eje $x_3$. En estos casos no se pueden excitar modos transversales horizontales (SH) y, por consiguiente, no se puede determinar la constante $C_{12}$. Sin embargo, al incluir la información de textura se observó empíricamente que esta indeterminación desaparece. Es decir, que para un material dado, no existirían dos texturas diferentes que resulten en tensores de rigidez, donde todas las constantes, salvo $C_{12}$, sean idénticas. Por ende, incluir la información de textura eliminaría la incertidumbre para este tipo de macro-simetría. Un análisis más profundo de esto se incluye en el apéndice \ref{app:incert_hex}.



\bibliographystyle{elsarticle-num} 
\bibliography{references.bib}


%%\bibitem{references.bib}

%%\end{thebibliography}
%}ending

\end{document}